\documentclass{article} 
\usepackage{times_style,times}
\usepackage{lipsum}
\usepackage{multirow}
\usepackage{booktabs}
\usepackage{multirow}
\usepackage{multicol}
\usepackage{adjustbox}
\usepackage{tcolorbox}
\usepackage{tabularx}
\usepackage{tikz}
\usepackage{graphicx}

\usepackage{amsmath,amsfonts,bm}

\def\eqref#1{equation~\ref{#1}}

\def\1{\bm{1}}

\DeclareMathAlphabet{\mathsfit}{\encodingdefault}{\sfdefault}{m}{sl}
\SetMathAlphabet{\mathsfit}{bold}{\encodingdefault}{\sfdefault}{bx}{n}

\usepackage{hyperref}
\usepackage{url}
\usepackage{lineno}
\usepackage{array, makecell}
\usepackage{algorithm}
\usepackage{algpseudocode}
\usepackage[numbers]{natbib}
\usepackage{geometry}
\usepackage{amssymb}
\usepackage{chngcntr}
\usepackage{subcaption}
\usepackage{enumitem}
\usepackage{overpic}
\usepackage{float} 
\title{\textsc{Pre-training Graph Neural Networks with Structural Fingerprints for Materials Discovery}}

\author{Shuyi Jia, Shitij Govil, Manav Ramprasad, Victor Fung$^\dag$ \\
School of Computational Science and Engineering\\
Georgia Institute of Technology\\
Atlanta, GA 30332, USA \\
$^\dag$Corresponding author: \texttt{victorfung@gatech.edu} \\
}

\date{}

\newcolumntype{Y}{>{\centering\arraybackslash}X}

\iclrfinalcopy 
\begin{document}

\maketitle

\begin{abstract}
\noindent In recent years, pre-trained graph neural networks (GNNs) have been developed as general models which can be effectively fine-tuned for various potential downstream tasks in materials science, and have shown significant improvements in accuracy and data efficiency. The most widely used pre-training methods currently involve either supervised training to fit a general force field or self-supervised training by denoising atomic structures equilibrium. Both methods require datasets generated from quantum mechanical calculations, which quickly become intractable when scaling to larger datasets. Here we propose a novel pre-training objective which instead uses cheaply-computed structural fingerprints as targets while maintaining comparable performance across a range of different structural descriptors. Our experiments show this approach can act as a general strategy for pre-training GNNs with application towards large scale foundational models for atomistic data. 
\end{abstract}

\section{Introduction}\label{sec:intro}

Graph neural networks (GNNs) have gained significant traction in materials science due to their adaptability and effectiveness across diverse applications \citep{fung2021benchmarking, reiser2022graph, chien2024opportunities}. Their rapid adoption is driven by the pressing challenges in materials discovery and evaluation, where conventional approaches for measuring or computing material properties are often costly, time-consuming, and reliant on complex synthesis and characterization techniques, or computationally demanding \textit{ab initio} simulations with density functional theory (DFT) \citep{curtarolo2013high, liu2017materials}. Furthermore, traditional machine learning models for property prediction typically require extensive feature engineering and domain expertise, limiting their generalizability across the material space \citep{musil2021physics}. In contrast, GNNs are powerful and broadly applicable models that capture intricate structural and chemical relationships with minimal feature engineering, enabling substantial advancements in property prediction \citep{gilmer2017neural, xie2018crystal, schutt2018schnet, satorras2021n, choudhary2021atomistic, yan2022periodic, yang2024mattersim}, materials screening \citep{xiong2019pushing, park2020developing, witman2023defect}, molecular dynamics simulations \citep{li2022graph, gasteiger2022gemnet}, and the inverse design of novel materials \citep{xie2018crystal, jiao2023crystal, zeni2025generative}.


A significant challenge for GNNs is their reliance on large amounts of training data, which are often scarce or difficult to obtain. A promising approach to address this limitation is pre-training, where models are initially trained on large upstream datasets before being fine-tuned on smaller downstream datasets \citep{erhan2010does, hendrycks2019using}.  This approach enables the learning of robust, transferable underlying representations, reducing the amount of task-specific data required for effective performance. Current pre-training strategies for GNNs in the materials domain can be broadly categorized into two classes: supervised and unsupervised. In supervised pre-training, GNNs are initially trained on large-scale datasets with explicit target properties, before being fine-tuned on the usually much smaller, task-specific datasets with related property objectives \citep{yamada2019predicting, gupta2021cross, shoghi2023molecules, jia2024derivative}. In contrast, unsupervised pre-training of GNNs eliminates the need for explicit labels by generating surrogate labels during the pre-training process \citep{hu2019strategies, liu2021pre}. This approach, also known as self-supervised learning, includes strategies such as pre-training via denoising, where the surrogate objective involves predicting the amount of noise added to the Cartesian coordinates of molecules or materials at equilibrium \citep{zaidi2022pre, neumann2024orb}.

An underlying limitation of both supervised and self-supervised pre-training strategies for GNNs is their dependence on computationally expensive quantum mechanical calculations to generated training data. In supervised pre-training, target labels are most commongly the DFT-calculated forces and energies. Additionally, variations in DFT settings across different datasets introduce distributional discrepancies, making it challenging to merge them into a larger, unified pre-training dataset \citep{bursch2022best}. Similarly, self-supervised pre-training with denoising still relies on DFT geometry optimization to obtain equilibrium structures, despite not requiring any specific labels. This suggests that while current pre-training strategies for GNNs are largely effective, they remain constrained by the availability of high-fidelity data and labels, preventing their scale-up to the levels seen in other domains such as image and text.


Motivated by the need for a pre-training strategy that can effectively utilize the extensive amount of unlabeled data, we propose pre-training GNNs using features derived from structural descriptors as the pre-training objective. Structural descriptors are physics-informed representations of molecular or materials structures, capturing essential information about their composition and geometry while respecting physical requirements such as symmetry, smoothness, and completeness \citep{musil2021physics}. For these reasons, structural descriptors such as ACSFs \citep{behler2007generalized} and SOAP \citep{bartok2013representing} have been widely used as machine learned force fields. In our study, we aim to leverage the characteristics of these descriptors as universally applicable physical priors for the GNN model. In the fine-tuning stage, the GNN is then free to further refine upon these priors towards the desired objective of interest. 


\section{Methodology}\label{sec:method}
\subsection{Background}
The structure of a periodic material can be succinctly captured by its unit cell—a minimal 3D arrangement of atoms that serves as the fundamental building block of the crystal. By repeating this unit cell infinitely in all directions, the full periodic crystal structure is reconstructed. Given a structure $S$ with $N$ atoms in its unit cell, it can be fully described by the tuple $S = \left(\mathbf{Z}, \mathbf{X}, \mathbf{L}  \right)$, where $\mathbf{Z} =\left(Z_1,\ldots,Z_N  \right) \in \mathbb{Z}^N_{+}$ is a vector of $N$ atomic numbers, $\mathbf{X} = \left(\boldsymbol{x}_1, \ldots, \boldsymbol{x}_N\right)\in \mathbb{R}^{N\times 3}$  represents the Cartesian coordinates of the $N$ atoms, and $\mathbf{L} = \left( \boldsymbol{l}_1,\boldsymbol{l}_2,\boldsymbol{l}_3 \right)\in\mathbb{R}^{3\times 3}$ is a matrix of lattice vectors. 

\subsection{Structural Descriptors}
In this work, we illustrate our approach by selecting three representative structural descriptors to generate pre-training labels during the pre-training process: 1) weighted atom-centered symmetry functions (wACSFs) \citep{gastegger2018wacsf}, 2) Gaussian multipoles (GMP) \citep{lei2022universal, lei2023gmp}, and 3) embedded atomic density (EAD) \citep{zhang2019embedded, yanxon2020pyxtal_ff}. All three descriptors are selected as featurizers that remain compact across the composition space, ensuring robustness to the elemental diversity of the pre-training dataset.

\subsubsection{Weighted Atom-centered Symmetry Functions}
Introduced as an invariant featurization method for high-dimensional neural network potentials \citep{behler2007generalized}, the atom-centered symmetry functions (ACSFs) are descriptors of a chemical system's geometry. In particular, the classic $G^2$ radial function of ACSFs is given by
\begin{equation}
    G^2_i = \sum_j \exp\left( -\eta \left( R_{ij} - R_s \right)^2 \right) \cdot f_c\left( R_{ij} \right), \label{eq:original-acsf}
\end{equation}
where $\eta$ and $R_s$ are parameters that determine the width and the shift away from the center of the Gaussians respectively. 
The interatomic distance between atoms $i$ and $j$ is represented as $R_{ij}$, and $f_c$ denotes a cutoff function, commonly the cosine cutoff function. However, a known problem of the original ACSFs is the rapid expansion of output feature vector dimensions as the number of distinct elements in the system increases. To address this challenge, wACSFs \citep{gastegger2018wacsf} offer a solution by implicitly incorporating the chemical environment's composition. This is achieved through the introduction of element-dependent weighting functions $g\left(Z_j\right)$ into Equation \ref{eq:original-acsf},
\begin{equation}
    G^2_i = g\left(Z_j\right)\sum_j \exp\left( -\eta \left( R_{ij} - R_s \right)^2 \right) \cdot f_c\left( R_{ij} \right) , \label{eq:wacsf}
\end{equation}
where $Z_j$ is the atomic number of atom $j$. In this work, we let $g\left(Z_j\right) = Z_j$.

\subsubsection{Gaussian Multipole}
The Gaussian multipole (GMP) featurization scheme extracts rotationally invariant feature vectors by leveraging multipole expansions of the electron density around atoms. These features interpolate between element types and are computed as inner products between the electron density and atom-centered \textit{probe} functions. The full GMP feature vector is defined as
\begin{align}
    \vec{\varphi} = \sqrt{w_{abc}\sum_{P(a,b,c)}\sum_i \mu^2_{i,abc}},
\end{align}
where $abc$ is the index of the spherical harmonic function with order $n=a+b+c$, $w_{abc} = n!/(a!b!c!)$, $P(a,b,c)$ is the set of all possible ordered combinations of $a$, $b$ and $c$, and
\begin{align}
    \mu_{i, abc} &= \left\langle S_{abc} \times G^{\text{probe}}_i, \sum_{j}\sum_{k} G^{\text{density}}_{jk} \right\rangle \\
    &= \sum_j \sum_k \iiint_V S_{abc} G^{\text{probe}}_i G^{\text{density}}_{jk} dV,
\end{align}
where $\langle\cdot,\cdot\rangle$ denotes the inner product of 2 functions, $V$ is the volume, $S_{abc}$ represents the angular probe and $G^{\text{probe}}$ is the radial probe. The electron density of the system is approximated by linear combinations of primitive Gaussians $G^{\text{density}}$ centered at each atom. For precise definitions of the probe functions and other additional details, please see Ref. \citep{lei2022universal}.

\subsubsection{Embedded Atom Density}
The embedded atom density (EAD) descriptor is inspired by the embedded atom method (EAM) \citep{zhang2019embedded, yanxon2020pyxtal_ff}, which assumes that each atom is embedded within the electron cloud of its neighboring atoms when modeling atomic bonding. 
EAD modifies EAM by including a linear combination of atomic orbital components:
\begin{equation}
    \rho_i(R_{ij}) = \sum_{l_x, l_y, l_z}^{L} \frac{L!}{l_x! \, l_y! \, l_z!} \left( \sum_{j \neq i} Z_j \Phi(R_{ij}) \right)^2, \label{eq:original-ead}
\end{equation}
where $Z_j$ represents the atomic number of neighboring atom $j$, $L$ represents the quantized angular momentum, and $l_x, l_y, l_z$ are the directional angular momentum components for the $x,y,z$ directions, respectively. The term $\Phi(R_{ij})$ is explicitly written as:
\begin{equation}
    \Phi(R_{ij}) = x_{ij}^{l_x} y_{ij}^{l_y} z_{ij}^{l_z} \cdot e^{-\eta (R_{ij} - R_s)^2} \cdot f_c(R_{ij}),
    \label{eq:ead-2}
\end{equation}
where $f_c(R_{ij})$, $\eta$ and $R_s$ are defined as in Eq. \ref{eq:original-acsf}. EAD improves upon Gaussian symmetry functions (e.g., ACSFs) by implicitly incorporating angular terms when $L>0$, whereas symmetry functions explicitly separate radial and angular components. This implicit encoding reduces computational cost by eliminating the need to sum over each neighbor individually.


\subsection{Graph Neural Networks}
The proposed pre-training strategy in this paper is model-agnostic and can be applied to any graph neural network (GNN) architecture for atomistic data. In this work, we perform our experiments with TorchMD-Net \citep{tholke2022torchmd}, a recently developed model based on an equivariant transformer architecture.

\subsection{Datasets}
Our descriptor-based pre-training dataset, referred to as ``MP Relaxed," consists of 62,783 equilibrium structures obtained from structural relaxations performed by the Materials Project \citep{chen2022universal}. For fine-tuning, we selected datasets targeting 8 properties from the MatBench suite \citep{dunn2020benchmarking}: exfoliation energy, frequency of the last phonon in the PhDOS peak, refractive index, shear modulus, bulk modulus, formation energy, and band gap. In addition to these, we incorporated three specialized datasets. The 2D materials dataset focuses on two-dimensional materials with work function as the target property \citep{haastrup2018computational}. The metal-organic frameworks (MOFs) dataset contains crystal structures where the target property is the band gap \citep{rosen2021machine}. Lastly, the metal alloy surfaces dataset (referred to as Surface) includes structures with adsorption energy as the target property \citep{mamun2019high}. An overview of these datasets is provided in Table~\ref{tab:ft-dataset}.

\begin{table}[ht]
\begin{center}
\begin{tabular}{rlrlr}
\toprule
& \multicolumn{1}{l}{\bf Dataset}  &\multicolumn{1}{c}{\bf \# Structures} & \multicolumn{1}{l}{\bf Property} & \multicolumn{1}{r}{\bf Unit}
\\ \midrule 
1.& JDFT       & 636         & Exfoliation energy 			&   meV/atom	\\
2.& Phonons    & 1,265        & Freq. at last phonon PhDOS peak	&   1/cm   \\
3.& Dielectric & 4,764        & Refractive index   			&   ---	\\
4.& (Log) GVRH       & 10,987       & Shear modulus      			&   GPa	\\
5.& (Log) KVRH       & 10,987       & Bulk modulus       			&   GPa	\\
6.& Perovskite & 18,928       & Formation energy   			&   eV/atom	\\
7.& MP Form    & 132,752      & Formation energy   			&   eV/atom	\\
8.& MP Gap     & 106,113      & Band gap           			&   eV	\\
9.& 2D         & 3,814        & Work function      			&   eV	\\
10.& MOF        & 13,058       & Band gap           			&   eV	\\
11.&    Surface    & 37,334       & Adsorption energy  			&   eV	\\ 
\bottomrule
\end{tabular}
\end{center}
\caption{Overview of the finetuning datasets used for benchmarking the
performance of the pre-trained models. The first eight datasets are part of the MatBench suite.}
\label{tab:ft-dataset}
\end{table}

\subsection{Training Setup}
The TorchMD-Net model is implemented in PyTorch as part of the MatDeepLearn package \citep{fung2021benchmarking}. We directly use the implementations of structural descriptors from StructRepGen \citep{fung2022atomic} and GMPFeaturizer \citep{lei2023gmp}. The finetuning experiments are averaged across 5 runs with different seeds. The \texttt{train:validation:test} split ratio is \texttt{0.6:0.2:0.2} for fine-tuning on every downstream dataset. Detailed hyperparameter settings can be found in Appendix \ref{appendix:implementation}.

\section{Results}
We begin by evaluating the effectiveness of our proposed pre-training strategy, which utilizes chemical descriptors as self-generated pre-training labels, on the 11 downstream datasets listed in Table \ref{tab:ft-dataset}. Specifically, we pre-train TorchMD-Net models independently using the three selected descriptors—wACSF, GMP, and EAD—for 200 epochs. To generate node-level features for each structure in the MP Relaxed dataset using wACSF, we set $R_s = [0,1,2,3,4,6,8,10]$ and $\eta = [0.01, 0.1, 0.4, 1.0, 3.5, 5.0]$, resulting in a feature dimension of 48. For GMP, we use $n = [-1,0,1,2]$ and $\sigma = [0.5, 1.0, 1.5, 2.0]$, yielding a feature dimension of 13. For EAD, we set $L=2$, $\eta = [1,5]$, and $R_s = [0.0:0.05:8.0]$, an arithmetic sequence from 0.0 to 8.0 with a common difference of 0.05, resulting in a feature dimension of 1920. After pre-training, we fine-tune each model on the downstream datasets for an additional 100 epochs. Notably, the pre-training descriptor labels are node-level, whereas the target properties in the downstream datasets are graph-level.

The finetuning results of pre-training with descriptors are summarized in Table \ref{tab:main-results}. Here, the baseline refers to training the models directly on the downstream datasets without any prior pre-training. From Table \ref{tab:main-results}, we observe that all three descriptors—wACSF, GMP, and EAD—demonstrate significant average improvements in MAE compared to the baseline. Notably, pre-training with EAD achieves the highest average MAE improvement of 16.9\% over the baseline, closely followed by wACSF at 16.8\%, while GMP ranks third with an average improvement of 15.2\%. Among the 11 datasets, EAD achieves the lowest MAE on 7, demonstrating its superior performance in most cases. Overall, the percentage improvement over the baseline for each individual dataset ranges from 4.97\% to 35.0\%, underscoring the effectiveness of pre-training the neural networks with descriptor labels. 

\begin{table}[h]
\centering
\resizebox{\textwidth}{!}{%
\setlength{\tabcolsep}{2.5pt} 
\renewcommand{\arraystretch}{1.3}
\begin{tabular}{@{}lccccccccccc|c@{}}
\toprule
& \textbf{JDFT} & \textbf{Phonons} & \textbf{Dielectric} & \textbf{GVRH} & \textbf{KVRH} & \textbf{Perovskites} & \textbf{2D} & \textbf{MOF} & \textbf{Surface} & \textbf{MP gap} & \textbf{MP form} & \textbf{Avg. \% Impr.} \\
\midrule
Baseline & 57.6 & 158.6 & 0.486 & 0.1100 & 0.0846 & 0.0478 & 0.298 & 0.254 & 0.0778 & 0.233 & 0.0354 & -- \\ \midrule
wACSF     & \textbf{44.0} & \textbf{106.5} & 0.404 & 0.0906 & 0.0630 & 0.0415 & 0.225 & 0.261 & 0.0662 & \textbf{0.219} & \textbf{0.0313} & 16.8 \\
GMP      & 45.2 & 111.8 & 0.385 & \textbf{0.0818} & 0.0593 & 0.0420 & 0.260 & 0.255 & 0.0743 & 0.234 & 0.0317 & 15.1 \\
EAD      & 49.6 & 135.0 & \textbf{0.348} & \textbf{0.0818} & \textbf{0.0550} & \textbf{0.0398} & \textbf{0.217} & \textbf{0.242} & \textbf{0.0653} & 0.230 & 0.0350 & \textbf{16.9} \\ \midrule
Best \% Impr.     & 23.7 & 32.9  & 28.4  & 25.9   & 35.0   & 16.8   & 27.3  & 4.97  & 16.2   & 6.02  & 11.6   & -- \\
\bottomrule
\end{tabular}%
}
\caption{MAEs on the finetuning datasets comparing the performance of TorchMD-Net pre-trained with different descriptors. Results are averaged over 5 runs, each with a different seed. Models are pre-trained for 200 epochs and finetuned for 100 epochs. The last column, avg. \% Impr., shows the percentage improvement in MAE averaged across a specific row. The lowest MAE for a specific dataset is highlighted in bold.}
\label{tab:main-results}
\end{table}

\begin{figure}[h]
     \centering
     \begin{subfigure}[b]{0.32\linewidth}
         \centering
         \includegraphics[width=\linewidth]{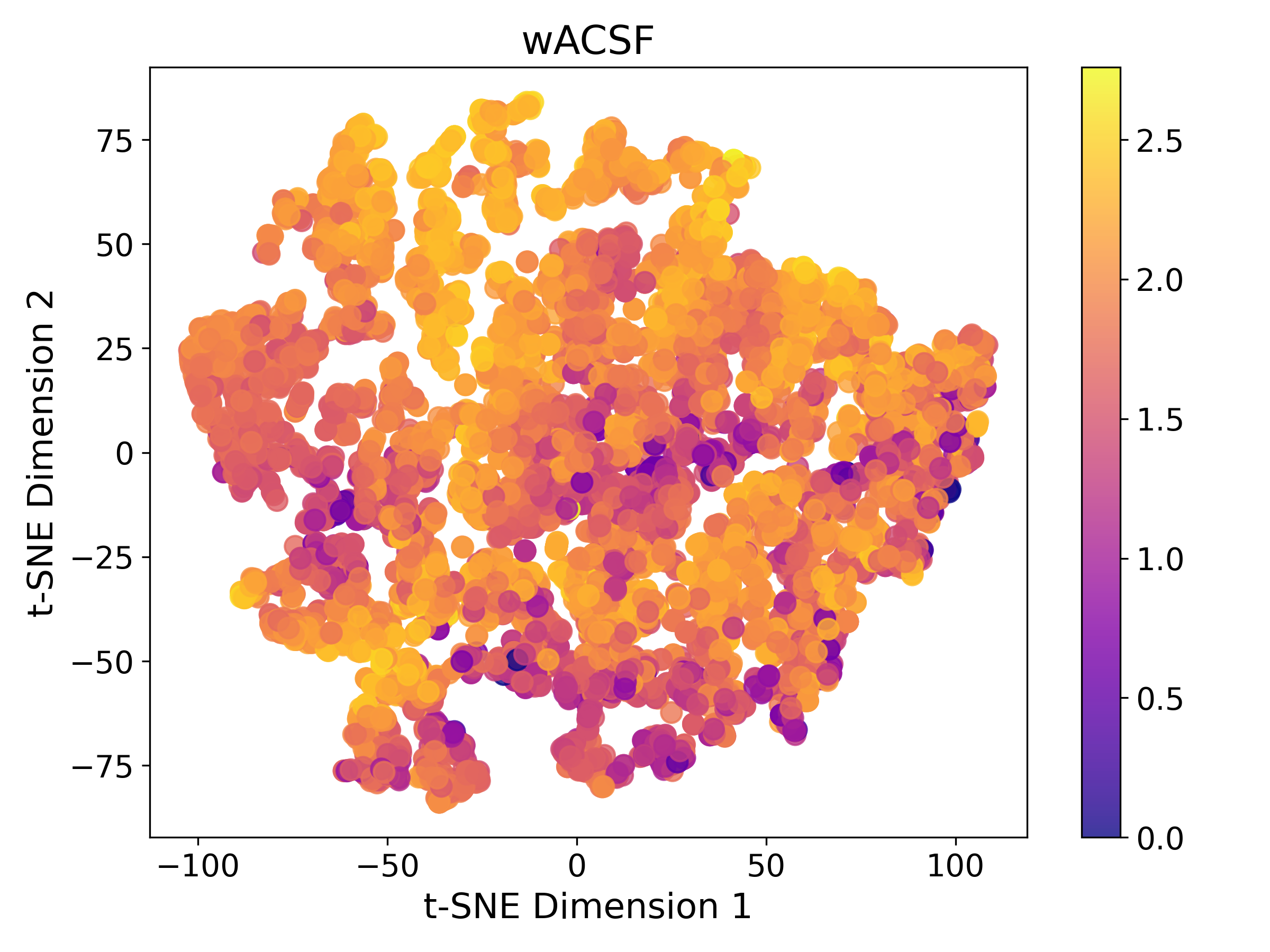}
         \caption{KVRH wACSF}
         \label{fig:kvrh-wacsf}
     \end{subfigure}
     \begin{subfigure}[b]{0.32\linewidth}
         \centering
         \includegraphics[width=\linewidth]{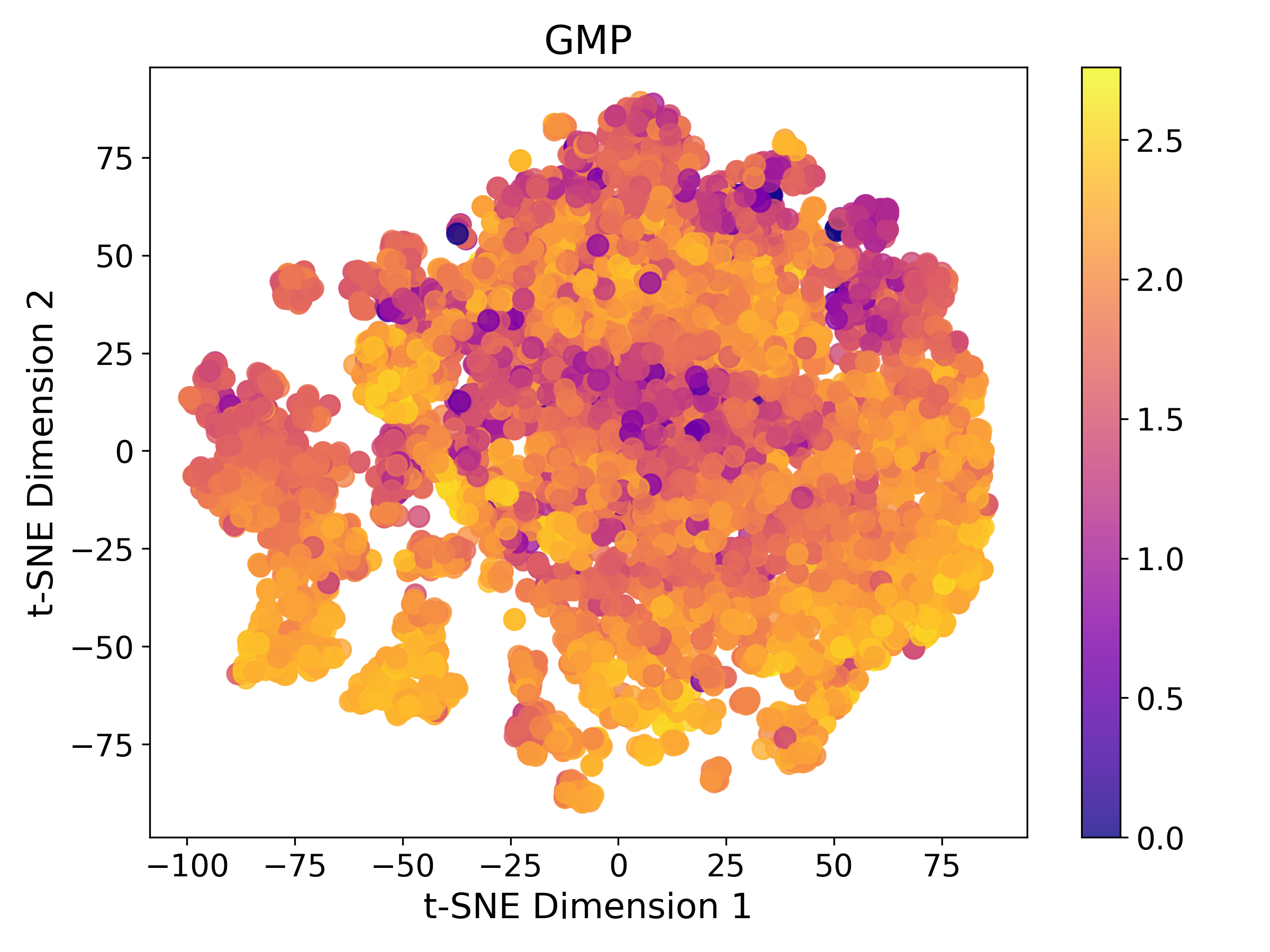}
         \caption{KVRH GMP}
         \label{fig:kvrh-gmp}
     \end{subfigure}
     \begin{subfigure}[b]{0.32\linewidth}
         \centering
         \includegraphics[width=\linewidth]{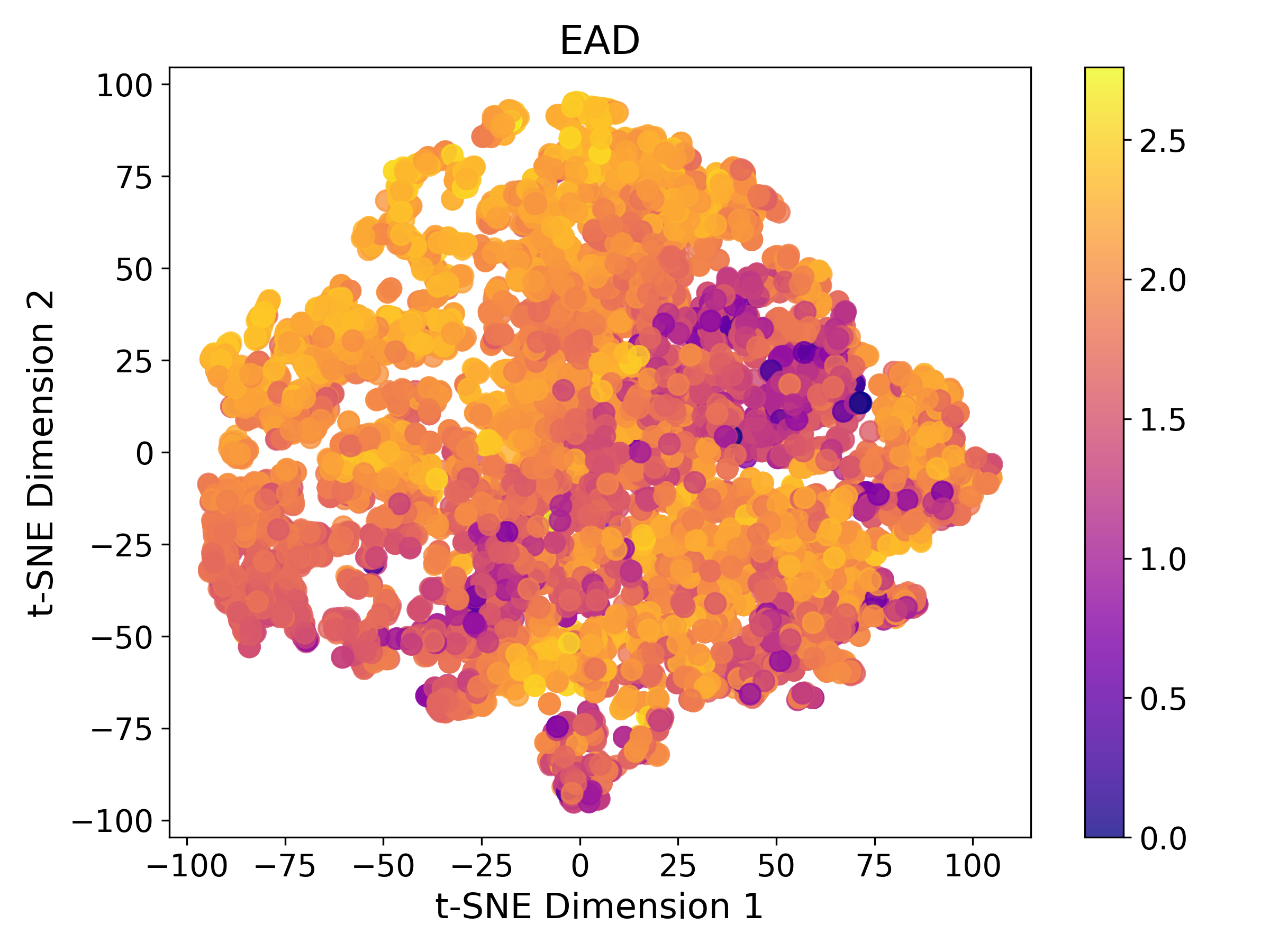}
         \caption{KVRH EAD}
         \label{fig:kvrh-ead}
     \end{subfigure}
     \begin{subfigure}[b]{0.32\linewidth}
         \centering
         \includegraphics[width=\linewidth]{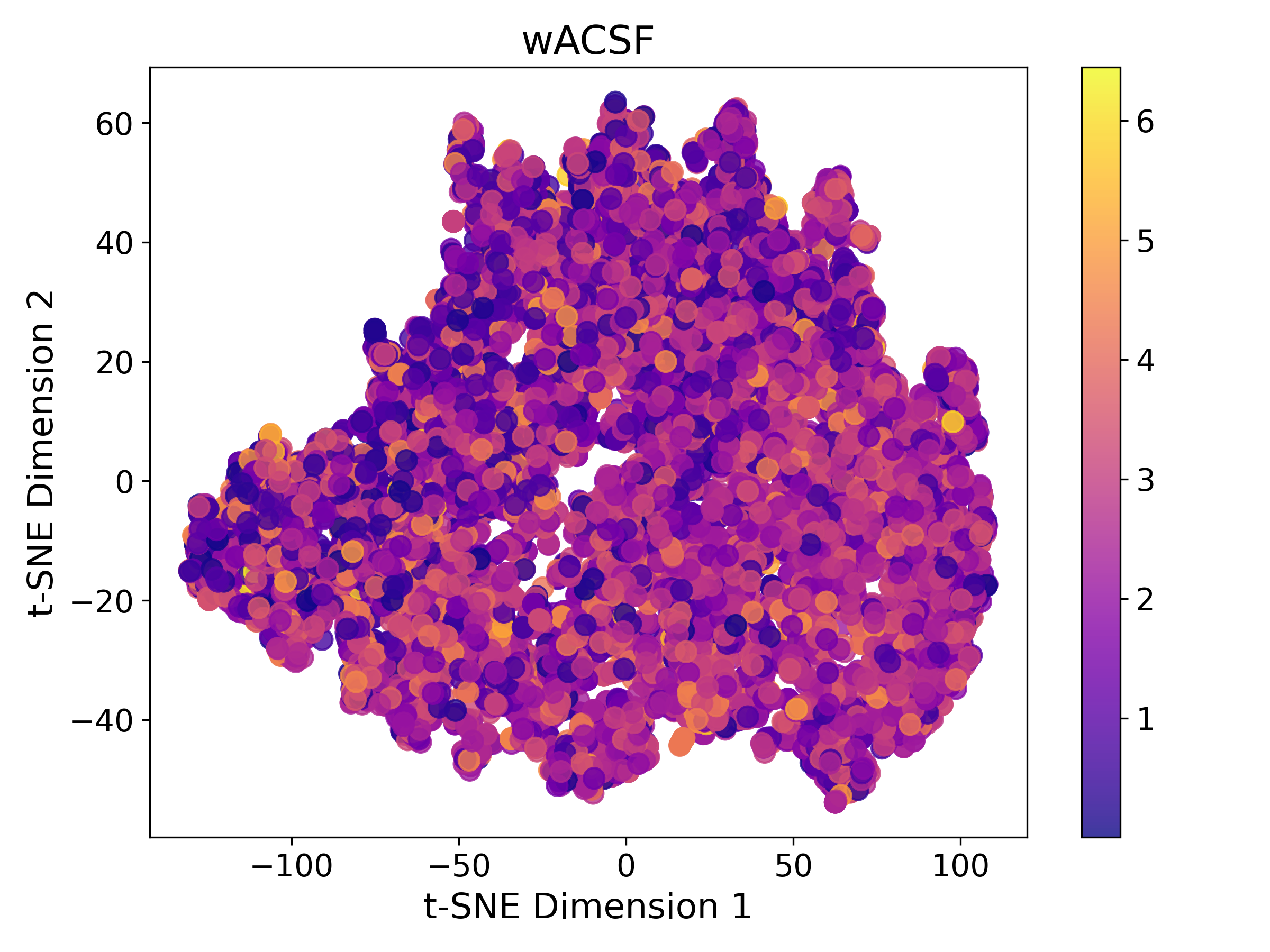}
         \caption{MOF wACSF}
         \label{fig:mof-wacsf}
     \end{subfigure}
     \begin{subfigure}[b]{0.32\linewidth}
         \centering
         \includegraphics[width=\linewidth]{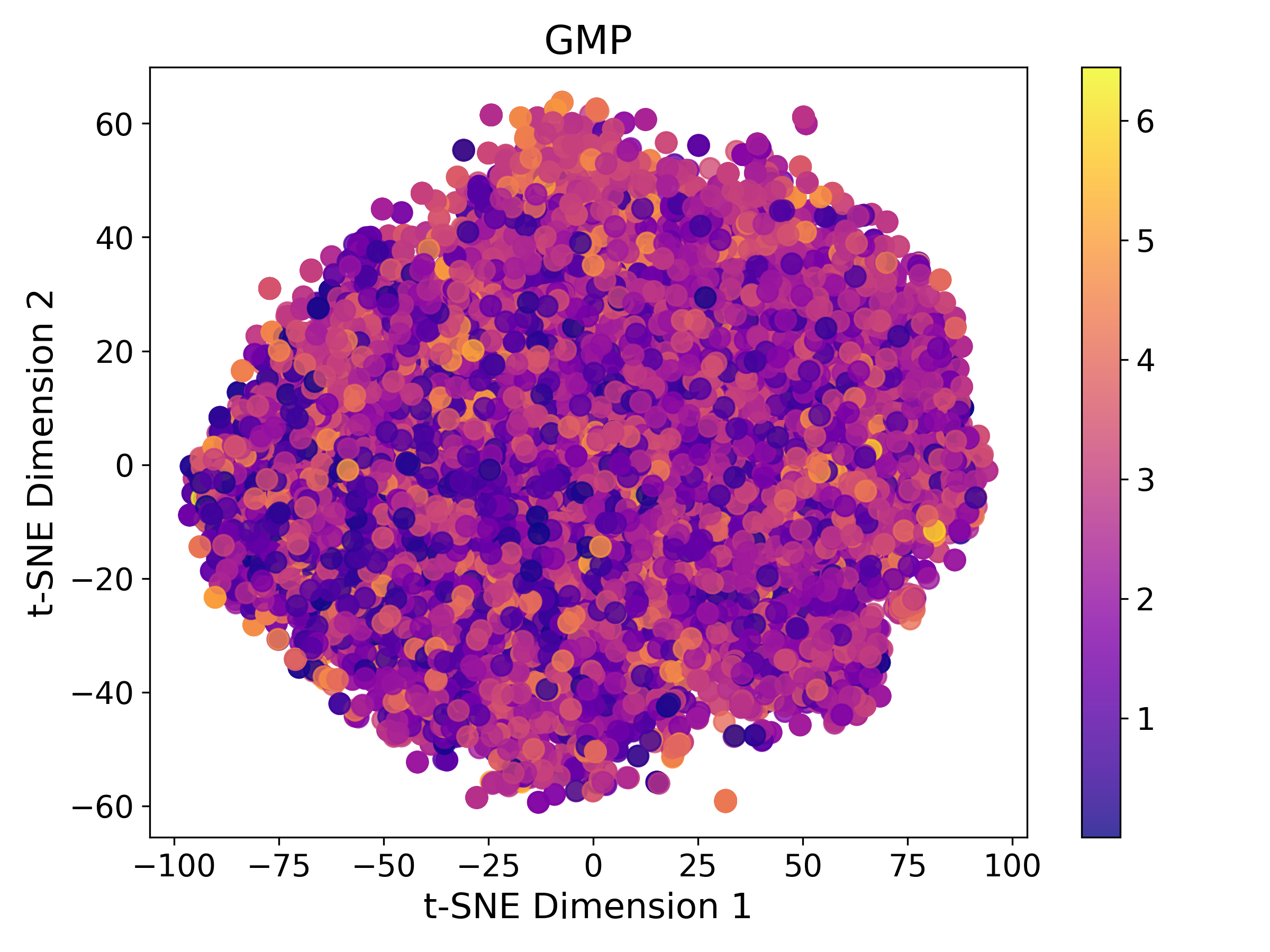}
         \caption{MOF GMP}
         \label{fig:mof-gmp}
     \end{subfigure}
     \begin{subfigure}[b]{0.32\linewidth}
         \centering
         \includegraphics[width=\linewidth]{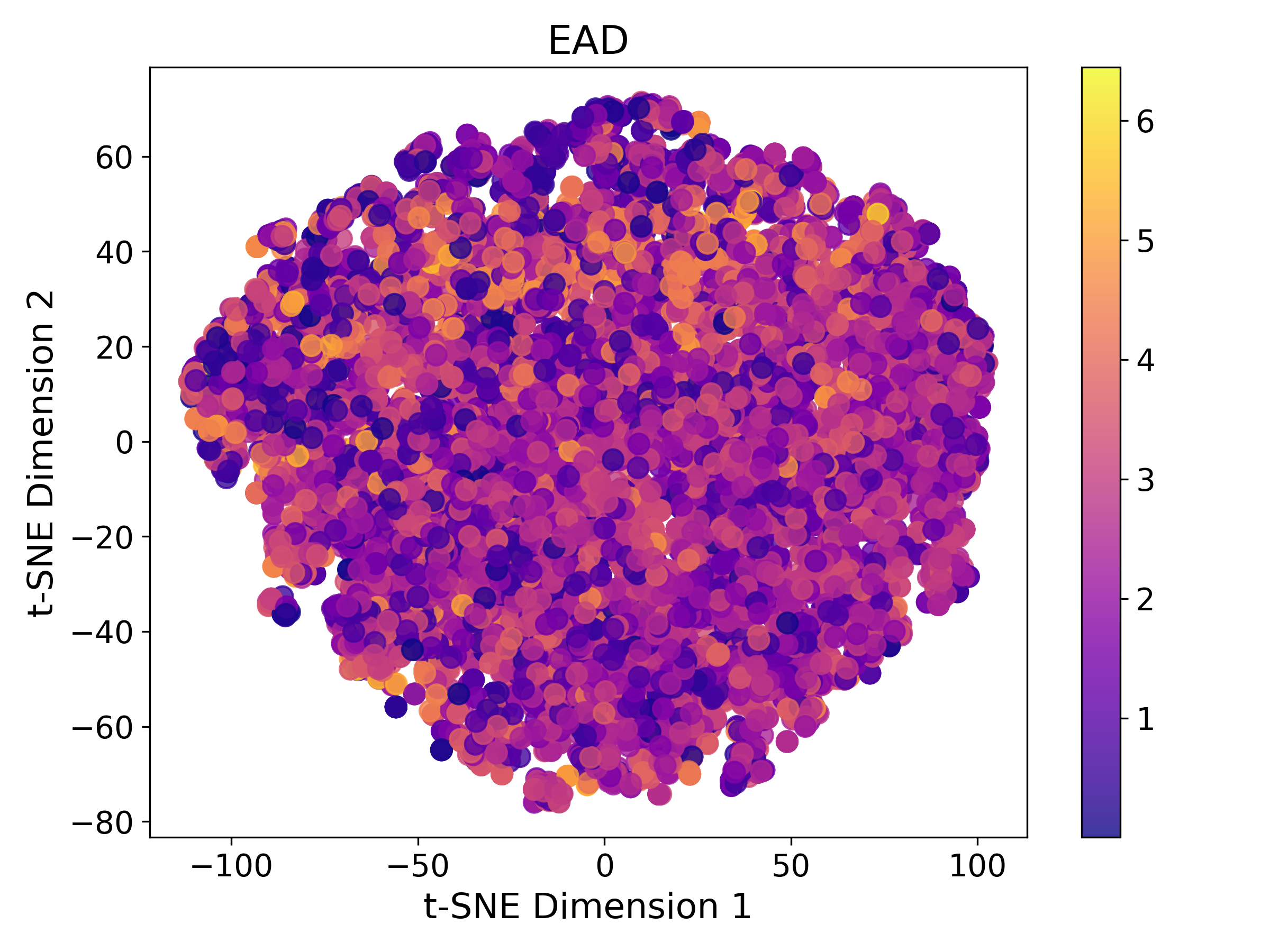}
         \caption{MOF EAD}
         \label{fig:mof-ead}
     \end{subfigure}
        \caption{Visualization of Graph-Level Embeddings: t-distributed stochastic neighbor embedding (t-SNE) plots of graph-level embeddings from the pooling layer for models pre-trained with wACSF, GMP, and EAD. Panels (a)-(c) correspond to the KVRH dataset; panels (d)-(f) correspond to the MOF dataset. Data points are color-coded based on bulk modulus (GPa) and band gap (eV), respectively.}
        \label{fig:embeddings_kvrh_mof}
\end{figure}

Notably, the three descriptors exhibit subpar performance on the MOF dataset, achieving only modest MAE improvements of less than 5\%. In particular, pre-training with wACSF fails to outperform the baseline on this dataset, yielding an MAE of 0.261 compared to the baseline's 0.254. To explore this behavior, we visualize the structural embeddings of the best-performing dataset, KVRH, and the worst-performing dataset, MOF, as shown in Fig. \ref{fig:embeddings_kvrh_mof}. To generate graph-level embeddings, each structure from the downstream datasets is first processed through the pre-trained TorchMD-Net models to obtain node-level embeddings. These node-level embeddings are subsequently aggregated using add-pooling to produce a single graph-level embedding for each structure. Finally, the high-dimensional graph-level embeddings are projected into a 2-dimensional space using t-distributed stochastic neighbor embedding (t-SNE). 

In Fig. \ref{fig:embeddings_kvrh_mof}, we compare the embeddings of TorchMD-Net models pre-trained with different chemical descriptors in capturing structural and compositional representations for the KVRH and MOF datasets. For the KVRH dataset, all three models pre-trained with the respective descriptors successfully generate meaningful representations, resulting in distinct clusters corresponding to structures with similar bulk modulus values. In contrast, the models face challenges in producing effective representations for the MOF dataset, where structures with similar band gap values fail to form clear groupings. This difference in the quality of graph-level embeddings highlights the disparity in fine-tuning performance between the two datasets. Specifically, the KVRH dataset achieves a substantial MAE improvement of up to 35.0\%, whereas the MOF dataset exhibits only a modest improvement of 4.97\%. This contrast can likely be attributed to the MOF dataset being more out-of-distribution relative to the pre-training dataset. Specifically, the MOF dataset includes much larger structures, averaging 84.8 atoms per structure, compared to 29.9 atoms in the pre-training dataset and just 8.63 atoms in the KVRH dataset. This size disparity likely hampers the model's ability to generalize effectively to the MOF dataset. 

\subsection{Descriptor Hyperparameter Tuning}
We investigate how different hyperparameter settings for the descriptors influence downstream fine-tuning performance. Specifically, for each descriptor, we introduce two additional hyperparameter sets alongside those used in Table \ref{tab:main-results}. We then pre-train models using these different hyperparameter sets and fine-tune them on the downstream datasets accordingly. The hyperparameter settings are shown in Table \ref{tab:params-sets}.

\begin{table}[H]
\centering
\resizebox{\textwidth}{!}{%
\setlength{\tabcolsep}{2.5pt} 
\renewcommand{\arraystretch}{1.3}
\begin{tabular}{@{}lccccccccccccc@{}}
\toprule
$\qquad \quad$& \multicolumn{3}{c}{\textbf{wACSF}} & \phantom{abc} & \multicolumn{3}{c}{\textbf{GMP}} & \phantom{abc} & \multicolumn{3}{c}{\textbf{EAD}} \\
\cmidrule{2-4} \cmidrule{6-8} \cmidrule{10-13}
& \footnotesize{$R_s$} & \footnotesize{$\eta$} & $d$ && \footnotesize{$n$} & \footnotesize{$\sigma$} & $d$ && \footnotesize{$L$} & \footnotesize{$R_s$} & $\eta$ & $d$ \\
\midrule
Set 1      & \makecell{$[0,1,2,3,$ \\ $4,6,8,10]$} & \makecell{$[0.01, 0.1, 0.4,$\\ $1.0, 3.5, 5.0]$} & 48 && $[-1,0,1,2]$  & $[0.5, 1.0, 1.5, 2.0]$ & 13 && $2$  & $[0.0:0.05:8.0]$ & $[1,5]$ & 1920 \\

Set 2 & $[0, 1, 2, 3, 4]$ & $[0.01, 0.1, 0.4, 1.0]$ & 20 && $[-1,0,1,2]$ & $[0.1, 0.2, 0.3]$ & 10 && $1$ & $[0.0:0.05:12]$ & $[1,2,5,10]$ & 3840 \\

Set 3 & \makecell{$[0, 1, 1.5, 2, 2.5, 3,$ \\ $3.5, 4, 5, 6, 7, 8, 9, 10]$} & \makecell{$[0.01, 0.06, 0.1, 0.2, 0.4,$ \\ $0.7, 1.0, 2.0, 3.5, 5.0]$} & 140 && $[-1, 0, 1, 2,3, 4]$ & \makecell{$[0.167, 0.333, 0.5,$\\ $0.667, 0.833, 1.0]$} & 31 && $1$ & $[0.0:0.05:8.0]$ & $[1,2,3,5]$ & 2560 \\
\bottomrule
\end{tabular}%
}
\caption{Different descriptor hyperparameter settings used to generate node-level features in $\mathbb{R}^{d}$. A cutoff radius of 8.0 is applied for Sets 1 and 3 of EAD, while Set 2 uses a cutoff radius of 12.0. Set 1 is used for the results obtained in Table \ref{tab:main-results}.}
\label{tab:params-sets}
\end{table}

\begin{table}[h]
\centering
\resizebox{\textwidth}{!}{%
\setlength{\tabcolsep}{2.5pt} 
\renewcommand{\arraystretch}{1.3}
\begin{tabular}{@{}lccccccccccc|c@{}}
\toprule
& \textbf{JDFT} & \textbf{Phonons} & \textbf{Dielectric} & \textbf{GVRH} & \textbf{KVRH} & \textbf{Perovskites} & \textbf{2D} & \textbf{MOF} & \textbf{Surface} & \textbf{MP gap} & \textbf{MP form} & \textbf{Avg. \% Impr.} \\
\midrule
Baseline 	& 57.6 & 158.6 & 0.486 & 0.1100 & 0.0846 & 0.0478 & 0.298 & 0.254 & 0.0778 & 0.233 & 0.0354 & -- 	\\ \midrule
wACSF-1 		& \textbf{44.0} & 106.5 & 0.404 & 0.0906 & 0.0630 & 0.0415 & 0.225 & 0.261 & 0.0662 & \textbf{0.219} & 0.0313 & 16.8  \\
wACSF-2		& 45.0 & 109.7 & 0.371 & 0.0895 & 0.0608 & 0.0420 & 0.234 & 0.257 & 0.0634 & 0.222 & \textbf{0.0303} & 17.6	\\
wACSF-3		& 44.5 & \textbf{103.9} & 0.373 & 0.0873 & 0.0593 & 0.0448 & 0.231 & 0.282 & 0.0671 & 0.226 & 0.0319 & 16.0  \\ \midrule
GMP-1			& 45.2 & 111.8 & 0.385 & 0.0818 & 0.0593 & 0.0420 & 0.260 & 0.255 & 0.0743 & 0.234 & 0.0317 & 15.1  \\
GMP-2		& 46.2 & 107.4 & 0.405 & 0.0815 & 0.0593 & 0.0422 & 0.260 & 0.254 & 0.0680 & 0.235 & 0.0316 & 15.5  \\
GMP-3		& 44.6 & 111.7 & 0.420 & \textbf{0.0812} & 0.0585 & 0.0420 & 0.261 & 0.254 & 0.0701 & 0.233 & 0.0317 & 15.2  \\ \midrule
EAD-1			& 49.6 & 135.0 & \textbf{0.348} & 0.0818 & \textbf{0.0550} & 0.0398 & 0.217 & 0.242 & 0.0653 & 0.230 & 0.0350 & 16.9  \\
EAD-2		& 45.7 & 107.1 & 0.387 & 0.0820 & 0.0570 & 0.0400 & 0.231 & \textbf{0.239}$^\dag$ & \textbf{0.0624} & 0.235 & 0.0353 & \textbf{17.8}  \\
EAD-3		& 52.4 & 128.7 & 0.369 & 0.0831 & 0.0556 & \textbf{0.0397} & \textbf{0.215} & 0.243 & 0.0633 & 0.233 & 0.0351 & 16.3  \\
\bottomrule
\end{tabular}%
}
\caption{MAEs on the finetuning datasets comparing the performance of TorchMD-Net pre-trained with different descriptors, evaluated under two additional hyperparameter configurations for each descriptor. The appended number denotes the specific hyperparameter set from Table \ref{tab:params-sets} that is used. Set 1 corresponds to the original settings used in Table \ref{tab:main-results}. Results are averaged over 5 runs, each with a different seed. Models are pre-trained for 200 epochs and finetuned for 100 epochs. $^\dag$ indicates that the value was obtained from a single run due to numerical instability in other runs.}
\label{tab:params-results}
\end{table}

Note that Set 1 in Table \ref{tab:params-sets} corresponds to the original settings used to obtain the results in Table \ref{tab:main-results}. The MAEs on the fine-tuning datasets for different hyperparameter settings are presented in Table \ref{tab:params-results}. The feature dimensions of each descriptor also vary significantly in magnitude: wACSF and GMP range from 20 to 140 and 10 to 31, respectively, whereas EAD spans from 1920 to 3940. This results in a difference of approximately 400-fold between the largest and smallest dimensions. As observed in Table \ref{tab:params-results}, despite variations in hyperparameter settings, the average percentage improvements in MAE over the baseline remain relatively stable. Notably, the maximum differences in average percentage improvement for wACSF, GMP, and EAD are 1.6\%, 0.4\%, and 1.5\%, respectively. This suggests that pre-training with descriptors is robust to moderate variations in descriptor parameter choices. Furthermore, the fact that EAD has up to 400 times more dimensions than GMP yet achieves only slightly better results implies that the effectiveness of pre-training with descriptors is not highly sensitive to feature dimensionality. The model appears capable of learning meaningful representations from each descriptor regardless of its dimensionality, at least within the tested range of hyperparameters. However this also suggests that a path towards improved pre-training performance may need to go beyond simple choices in the hyperparameters.

\subsection{Pre-training Epoch Ablation}
\begin{figure}[h]
    \centering
    \includegraphics[width=\linewidth]{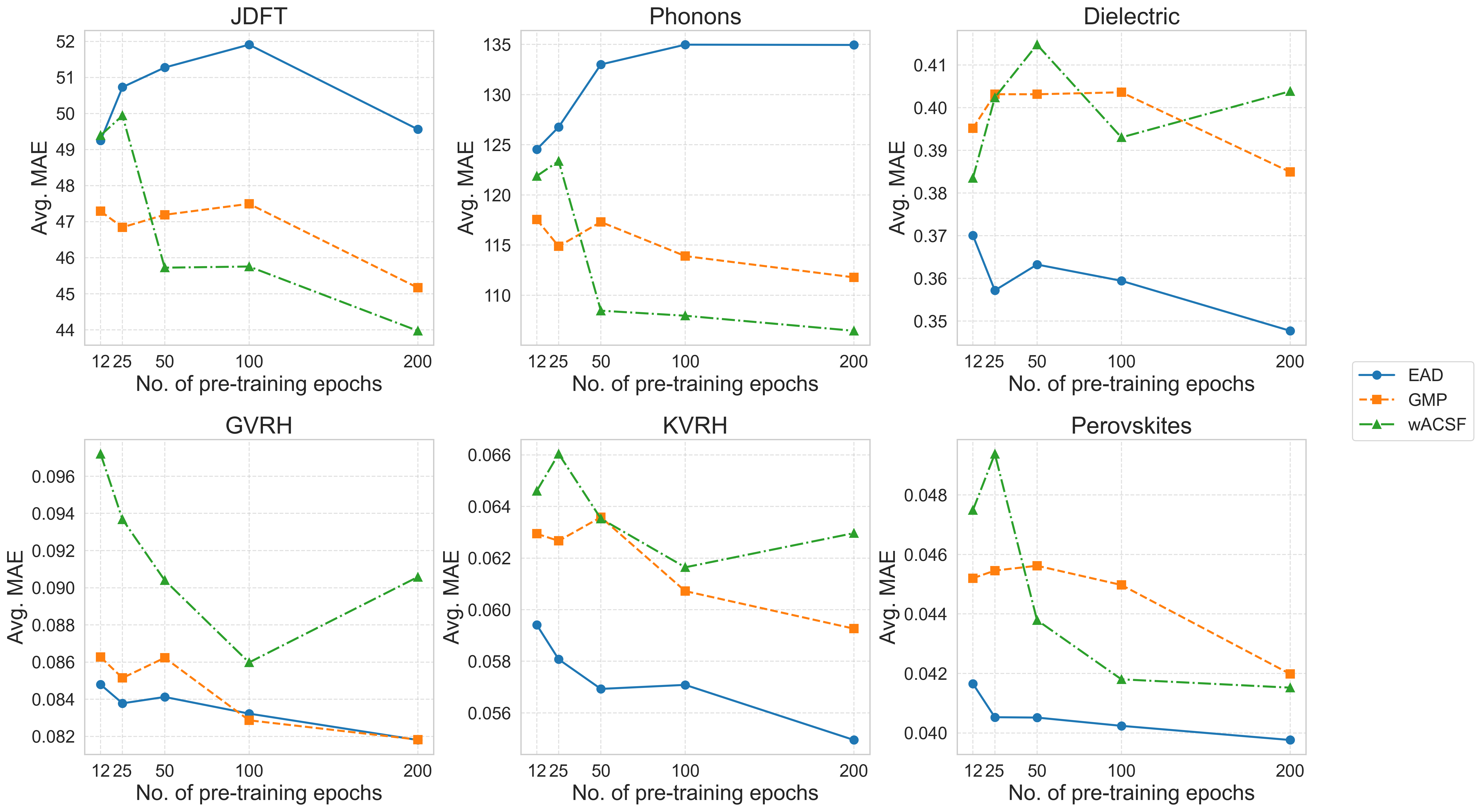}
    \caption{MAEs on the fine-tuning datasets for models pre-trained at 12, 25, 50, 100, and 200 epochs. The results corresponding to 200 epochs are from Table \ref{tab:main-results}.}
    \label{fig:ablation_trend}
\end{figure}

We investigate how the number of pre-training epochs on descriptor labels influences downstream test performance during fine-tuning. The results shown in Table \ref{tab:main-results} are based on models pre-trained for 200 epochs. In this section, we further explore four additional pre-training durations---12, 25, 50, and 100 epochs---for all 3 structural descriptors. We fine-tune only the first six datasets to save computational resources while still capturing the overall trend of pre-training epochs on downstream performance. Fig. \ref{fig:ablation_trend} shows the MAEs on downstream datasets plotted against the number of pre-training epochs. Overall, we observe a general trend where increasing the number of pre-training epochs leads to a reduction in average MAEs. This trend is particularly evident for the GVRH, KVRH, and perovskites datasets. For the EAD descriptor, a decreasing trend is observed across most datasets; however, this trend is not observed for the JDFT and phonons datasets, likely due to their smaller sizes and the presence of outliers. In the case of wACSF, while the general trend holds, the results suggest that pre-training the model for 100 epochs may be sufficient for certain datasets. The detailed MAEs on the fine-tuning datasets for different number of pre-training epochs are shown in Appendix \ref{appendix:epoch-ablation}.

\section{Discussion and Conclusion}
Our experiments demonstrate a significant improvement in performance using GNNs pre-trained on structural descriptors, observed consistently across all three tested descriptors. While the improvement is slightly lower on average than pre-training with forces \citep{jia2024derivative}, it remains competitive across the evaluated datasets. This slight reduction in performance is outweighed by the drastically lower cost of generating labels for this approach, reducing computational expense by several orders of magnitude. Within the class of pre-training approaches which do not require any quantum mechanical datasets or labels, this method yields state-of-the-art performance to the best of our knowledge. 

Therefore, this approach holds significant promise as a pre-training strategy for foundational GNNs which could be scaled up to billions of training samples or more on appropriately sized GNN models. Subsequent work is required to accomplish this, including the development and optimization of more effective physics-informed structural descriptors, as well as the creation of large-scale pre-training datasets that comprehensively span the materials space. Additionally, we anticipate that this approach can be integrated with existing pre-training strategies on quantum mechanical datasets, leveraging the strengths of both methodologies for enhanced performance.

\section{Data Availability}
 All data and materials used in this study are publicly accessible. The code supporting this work is available on GitHub at the following repository: \url{https://github.com/Fung-Lab/MatDeepLearn_dev}.

\section{Acknowledgements}

This research used resources of the National Energy Research Scientific Computing Center (NERSC), a U.S. Department of Energy Office of Science User Facility located at Lawrence Berkeley National Laboratory, operated under Contract No. DE-AC02-05CH11231 using NERSC award BES-ERCAP0022842.




\clearpage
\bibliography{refs}
\bibliographystyle{naturemag}
\clearpage

\appendix
\begin{titlepage}
  \centering
  \LARGE \textsc{Supplementary Information} \par
  \let\endtitlepage\relax
\end{titlepage}

\counterwithin{figure}{section}
\counterwithin{table}{section}

\section{Implementation Details}\label{appendix:implementation}
Our implementation of the TorchMD-Net model is based on the MatDeepLearn package \citep{fung2021benchmarking}. Graph representations are calculated through the algorithms released by the Open Catalyst Project (OCP) \citep{ocp_dataset}.

The main hyperparameters for TorchMD-Net are shown in Table \ref{tab:params-torchmdnet}.

\begin{table}[ht]
\caption{TorchMD-Net hyperparameters for pre-training with descriptors.}
\label{tab:params-torchmdnet}
\centering
\begin{tabular}{lr}
\toprule
Parameter                        & Value or function \\ [2pt]\midrule
Hidden channels                  & 128                  \\
Number of filters                & 128                  \\
Number of layers                 & 8                    \\
Number of RBF                    & 50                    \\
RBF type & expnorm                    \\
Trainable RBF                    & True                 \\
Cutoff distance & 8.0   \\
Max number of neighbors & 32 \\
Activation                       & silu                    \\
Attention activation              & silu                    \\
Number of heads                  & 8                    \\
Distance influence                     & both                    \\
Number of post layers                     & 2                    \\
Post hidden channels & 128 \\
Pooling                          & Global add pool      \\
Learning rate & 0.0001 \\
Batch size & 8 \\
Loss type & L1 loss \\
Optimizer & AdamW \\
\bottomrule
\end{tabular}
\end{table}

The pre-training dataset, namely MP Relaxed, is split on a \texttt{train:test:val} ratio of \texttt{0.8:0.15:0.05}. All finetuning datasets are split on a \texttt{train:test:val} ratio of \texttt{0.6:0.2:0.2} to ensure consistency and fair comparison. Models are pre-trained for 200 epochs and finetuned for 100 epochs for all datasets.
\clearpage
\section{Pre-training Epoch Ablation}\label{appendix:epoch-ablation}
\begin{table}[H]
\centering
\resizebox{0.7\textwidth}{!}{%
\setlength{\tabcolsep}{3.5pt} 
\renewcommand{\arraystretch}{1.3}
\begin{tabular}{@{}lcccccc|c@{}}
\multicolumn{8}{@{}l}{\hspace{0pt}\textbf{wACSF}} \\
\toprule
\# Pre-training Epoch & \textbf{JDFT} & \textbf{Phonons} & \textbf{Dielectric} & \textbf{GVRH} & \textbf{KVRH} & \textbf{Perovskites} & \textbf{Avg. \% Impr.} \\
\midrule
Baseline   & 57.6 & 158.6 & 0.486 & 0.1100 & 0.0846 & 0.0478 & -- \\ \midrule
12& 49.4   & 121.9 & \textbf{0.383} & 0.0972 & 0.0646 & 0.0475 & 15.8 \\
25& 49.9   & 123.4 & 0.402 & 0.0937 & 0.0660 & 0.0494 & 14.4 \\
50& 45.7   & 108.5 & 0.415 & 0.0904 & 0.0635 & 0.0438 & 19.7 \\
100& 45.8  & 108.0 & 0.393 & \textbf{0.0860} & \textbf{0.0616} & 0.0418 & \textbf{22.2} \\
200& \textbf{44.0}  & \textbf{106.5} & 0.404 & 0.0906 & 0.0630 & \textbf{0.0415} & 21.7 \\ \midrule
Best \% Impr. & 23.7 & 32.9  & 21.1  & 22.1   & 27.1   & 13.1  & -- \\
\bottomrule
\end{tabular}%
}
\caption{MAEs on the fine-tuning datasets comparing the impact of different pre-training epoch durations on fine-tuning performance. The models are pre-trained using the wACSF descriptor for 12, 25, 50, 100, and 200 epochs, respectively, and are then fine-tuned for 100 epochs.}
\label{tab:epoch-ablation-acsf}
\end{table}

\begin{table}[H]
\centering
\resizebox{0.7\textwidth}{!}{%
\setlength{\tabcolsep}{3.5pt} 
\renewcommand{\arraystretch}{1.3}
\begin{tabular}{@{}lcccccc|c@{}}
\multicolumn{8}{@{}l}{\hspace{0pt}\textbf{GMP}} \\
\toprule
\# Pre-training Epoch & \textbf{JDFT} & \textbf{Phonons} & \textbf{Dielectric} & \textbf{GVRH} & \textbf{KVRH} & \textbf{Perovskites} & \textbf{Avg. \% Impr.} \\
\midrule
Baseline   & 57.6 & 158.6 & 0.486 & 0.1100 & 0.0846 & 0.0478 & -- \\ \midrule
12 & 47.3 & 117.5 & 0.395 & 0.0863 & 0.0629 & 0.0452 	& 19.2 \\
25 & 46.8 & 114.9 & 0.403 & 0.0851 & 0.0627 & 0.0455	& 19.5 \\
50 & 47.2 & 117.3 & 0.403 & 0.0862 & 0.0636 & 0.0456	& 18.7 \\
100& 47.5 & 113.9 & 0.404 & 0.0829 & 0.0607 & 0.0450  	& 20.3   \\
200& \textbf{45.2} & \textbf{111.8} & \textbf{0.385} & \textbf{0.0818} & \textbf{0.0593} & \textbf{0.0420}	& \textbf{23.3} \\ \midrule
Best \% Impr. & 21.6 & 29.5  & 20.8  & 25.9   & 29.9   & 12.2  & -- \\
\bottomrule
\end{tabular}%
}
\caption{MAEs on the fine-tuning datasets comparing the impact of different pre-training epoch durations on fine-tuning performance. The models are pre-trained using the GMP descriptor for 12, 25, 50, 100, and 200 epochs, respectively, and are then fine-tuned for 100 epochs.}
\label{tab:epoch-ablation-gmp}
\end{table}

\begin{table}[H]
\centering
\resizebox{0.7\textwidth}{!}{%
\setlength{\tabcolsep}{3.5pt} 
\renewcommand{\arraystretch}{1.3}
\begin{tabular}{@{}lcccccc|c@{}}
\multicolumn{8}{@{}l}{\hspace{0pt}\textbf{EAD}} \\
\toprule
\# Pre-training Epoch & \textbf{JDFT} & \textbf{Phonons} & \textbf{Dielectric} & \textbf{GVRH} & \textbf{KVRH} & \textbf{Perovskites} & \textbf{Avg. \% Impr.} \\
\midrule
Baseline   & 57.6 & 158.6 & 0.486 & 0.1100 & 0.0846 & 0.0478 & -- \\ \midrule
12&  \textbf{49.2} & \textbf{124.5} & 0.370 & 0.0848 & 0.0594 & 0.0417 & 20.9\\
25&  50.7 & 126.8 & 0.357 & 0.0838 & 0.0581 & 0.0405 & 21.5\\
50&  51.3 & 133.0 & 0.363 & 0.0841 & 0.0569 & 0.0405 & 20.7\\
100& 51.9 & 135.0 & 0.359 & 0.0832 & 0.0571 & 0.0402 & 20.6\\
200& 49.6 & 135.0 & \textbf{0.348} & \textbf{0.0818} & \textbf{0.0550} & \textbf{0.0398} & \textbf{22.5}\\ \midrule
Best \% Impr. & 14.5 & 21.4  & 28.4  & 25.9   & 35.0   & 16.8  & -- \\
\bottomrule
\end{tabular}%
}
\caption{MAEs on the fine-tuning datasets comparing the impact of different pre-training epoch durations on fine-tuning performance. The models are pre-trained using the EAD descriptor for 12, 25, 50, 100, and 200 epochs, respectively, and are then fine-tuned for 100 epochs.}
\label{tab:epoch-ablation-ead}
\end{table}

\end{document}